\documentclass{elsart}

\usepackage{natbib}

\usepackage{graphicx}

\usepackage{amssymb}

\newcommand{\ssst}{\scriptscriptstyle}

\newcommand{\RA}[3]{{#1}^{{\rm h}}{#2}^{{\rm m}}{#3}^{{\rm s}}}
\newcommand{\Dec}[3]{{#1}^{\circ}{#2}'{#3}''}

\newcommand{\s}{\,{\rm s}}      \newcommand{\ps}{\,{\rm s}^{-1}}
    
\newcommand{\cm}{\,{\rm cm}}    

\newcommand{\ergs}{\,{\rm ergs}}        
    \newcommand{\keV}{\,{\rm keV}}

        \newcommand{\NH}{N_{\ssst\rm H}}

\newcommand{\nH}{n_{\ssst\rm H}}

\begin{document}

\begin{frontmatter}



\title{A Preliminary {\sl Chandra} X-ray Spectroscopy of the Supernova Remnant N132D}

 \author{Xiao Xiao},
 \author{Yang Chen}
 \address{Department of
Astronomy, Nanjing University, Nanjing 210093,
       P.R.China}

\begin{abstract}
We present the preliminary results of a Chandra X-ray study of
N132D, a young shell-like supernova remnant (SNR) in the Large Magellanic
Cloud. The equivalent width maps of emissions from O, Ne, Mg, Si,
and S are provided. Spatially resolved spectral analysis for the
small-scale regions were tentatively performed. The X-ray spectra of the
interior can be described with a single-thermal model. The faint
interior regions have lower density and higher temperature (above 1keV)
than those of bright interior regions. The X-ray spectra along the
shell can be phenomenally fitted with either a double-{\em vpshock}
model or a {\em vpshock} + {\em powerlaw} model. If the non-thermal
component is true, N132D would be listed as another X-ray
synchrotron SNR.

\end{abstract}


\end{frontmatter}


\section{Introduction}
 N132D, one of the prototype O-rich supernova remnants (SNRs) of a massive
progenitor, is located in the bar of the Large Magellanic Cloud
(LMC) (Danziger \& Dennefeld, 1976; Lasker, 1978). In X-ray
morphology, N132D is an irregular shell of radius about $50''$ with
a break-out in the northeast (Mathewson et al., 1983). N132D is a
good object to test the stellar evolution and nucleosynthesis models
of massive stars since the interior fragments are exposed to direct
investigation (Blair et al., 2000). N132D has two advantages for
observation. First, because it is located in the LMC, the distance
is well determined (50kpc) and the extinction is low. Second, N132D
is one of the brightest soft X-ray sources in the LMC (Favata et
al., 1997).  Multi-band observations have been performed on it.

In X-rays, Hwang et al. (1993) obtained relatively high resolution
spectrum of N132D with the Einstein Solid State Spectrometer.
Emission lines of $O^{6+}$, $O^{7+}$, $Ne^{9+}$, and $Fe^{16+}$ were
detected. In their work, the best-fit element abundances derived
with a single-temperature non-equilibrium ionization (NEI) model are
lower than the LMC mean abundances. With the data obtained by the
concentrator spectrometer on board the X-ray satellite BeppoSAX,
Favata et al. (1997, 1998) found that there should be two components
with temperatures of 3.3keV and 0.79keV respectively. The element
abundances derived with the two thermal components model are similar
to the normal LMC abundances. Hughes et al.\ (1998) first reported
the CCD data of N132D obtained by the Solid State Imaging
Spectrometer on board the Advanced Satellite for Cosmology and
Astrophysics (ASCA). They fitted the ASCA data of high resolution
with a single Sedov model and found that the temperature is around
0.7keV and the element abundances are all around the LMC mean, too.
Based on the ASCA observation, Chen et al.\ (2003) modeled the remnant
evolution as the blast wave hitting the pre-existing cavity wall.
With the extremely high spectral resolution of the Reflection
Grating Spectrometers on board the XMM-Newton X-ray observatory,
Behar et al. (2001) detected lines of C, N, O, Ne, Mg, Si, S, Ar,
Ca, and Fe in the spectrum of N132D. Images in the narrow
wave-length bands show that, with the exception of O, the dominant
part of the soft X-ray emission originates from shocked interstellar
medium (ISM). Using High Energy Transmission Grating (HETG) on board
Chandra, Canizares et al.\ (2001) got dispersed high resolution
X-ray images of N132D in which some regions of oxygen-rich material
were clearly identified despite the spatial/spectral overlap.

In the optical and UV bands, with IUE data, Blair et al. (1994)
noted that the X-ray peaks near the center were generally associated
with low-velocity, normal composition, optical filaments, rather
than with high-velocity, oxygen-rich ones. Morse et al. (1996)
obtained high spatial resolution images of N132D with Hubble Space
Telescope (HST). They distinguished oxygen-rich filaments from
shocked clouds. Comparing with the ROSAT images, they confirmed that
the O-rich ejecta emit little X-rays. Blair et al. (2000) performed
detailed spectral analysis on the data from HST. The abundance they
derived from the shocked ISM cloud in N132D is consistent with the
LMC mean. From the spectra of the O-rich ejecta they clearly
detected O, C, Ne, and Mg lines, but found no evidence of O-burning
elements. They suggested that N132D is the remnant of a Type Ib
supernova explosion of a W/O progenitor star with extensive O-rich
mantle.

In radio, Dickel \& Milne (1995) found that the 6 $cm$ radio
emission from N132D largely coincide with the X-ray shell, in other
words the shell of N132D seems to be the source of both X-ray and
synchrotron radio emission.

Recently, Tappe et al. (2006) detected strong 24 $\mu m$ infrared
emission from swept-up, shock-heated dust grains in the N132D with
Spitzer Space Telescope. The image of 24 $\mu m$ emission follows
the X-ray image well. Additionally, they detected PAH molecular
bands from N132D.

Profited from the high spatial resolution of Chandra, we have
performed a spatially resolved spectroscopic analysis of N132D and
present a fresh look of the physical properties of the remnant.

\section{Observation and Data Analysis}

We combined two sets of data (ObsId 121 and 1821 observed by C.R.
Canizares) released by Chandra X-ray center. Both of the data were
dispersed by the HETG at first, then read out by the Advanced CCD
Imaging Spectrometer (ACIS). The two observations were carried out
on 2000 July 19 with exposure time of 22ks and on 2000 July 20 with
exposure time of 74 ks, respectively.

The tool we used to process the data is Chandra Interactive Analysis
of Observations (CIAO) software package (ver. 3.2). Considering the
effect of the spatial/spectral overlap after dispersion by the HETG,
we only analyzed the zeroth order data. The two sets of data were
reprocessed separately to generate level 2 event files following the
threads for extended sources (the threads are available on
http://cxc.harvard.edu/ciao/). During the course of the
reprocessing, we corrected hot pixels and cosmic ray afterglows,
filtered bad grades and applied good time intervals correction. Then
we merged the two sets of cleaned data for subsequent analysis.

\subsection{Spatial Analysis} Fig.1 shows diffuse emission from SNR
N132D in the broad band 0.3-8.0 keV. Regions shown in the map are
used for spectral analysis (see section 2.2). This image displays a
horseshoe morphology with a bright ridge south to the center,
similar to that found in earlier X-ray observations (Mathewson et
al., 1983; Hughes, 1987; Behar et al., 2001) as well as the recent
IR observation (Tappe et al., 2006). It is easy to distinguish a
bright shell with a break-out in the northeast. Noticeably, there
are some bright knots and thin filaments along the shell.

The tricolor X-ray image is shown in Fig.2. The 0.3-1.0keV emission
is coded in red, 1.0-2.0keV in green, and 2.0-8.0keV in blue. The
images in the three bands were adaptively smoothed using CIAO tool
{\em csmooth} with a signal-to-noise ratio of 3 respectively. In the
tricolor image, soft emission (in red) is basically observed all
over the image. Soft emission dominates the northeast region where
the break-out locates. Southwestward, the proportion of hard
emission (in blue) increases. Hard emission could be easily
distinguished in the south part of the horseshoe-like shell.

From the overall spectrum of N132D (Fig.3), we can distinguish
emission lines of element species O, Ne, Mg, Si, S, Ar, Ca, and Fe,
as are seen in the XMM spectrum (Behar et al., 2001). Considering
the low signal-to-noise ratio in the high energy band, we only
present the equivalent width (EW) maps of O, Ne, Mg, Si, and S. Here
we have applied an adaptive mesh method to rebin the data so as to
include at least 10 counts in each bin. EW map indicates regions of
high line to continuum ratio, which varies with element abundances,
temperature, ionization parameter, and even column density (Hwang et
al., 2000).

The EW of O (left panel of Fig.4) is high in the north near the
breakout. In this map, we also see a shell which follows the X-ray
intensity shell but is thinner than it. The EW  of Ne (right panel
of Fig.4) is comparatively uniform across the SNR except for some
bright knots along/near the shell. The
strength distribution of Mg EW (left panel of Fig.5) follows the
X-ray intensity image well but it seems to have smaller extent than
that of O and Ne, especially in the north. The EW of Si (right panel
of Fig. 5) is fairly uniform, but the EW in an interior patch around
($\RA{05}{25}{04}$, $\Dec{-69}{38}{24}$) is roughly twice as high as
the average.  The EW of S (Fig. 6) is enhanced in the
intensity shell and an interior patch near the center.

\begin{figure}
\begin{center}
\includegraphics[scale=0.7]{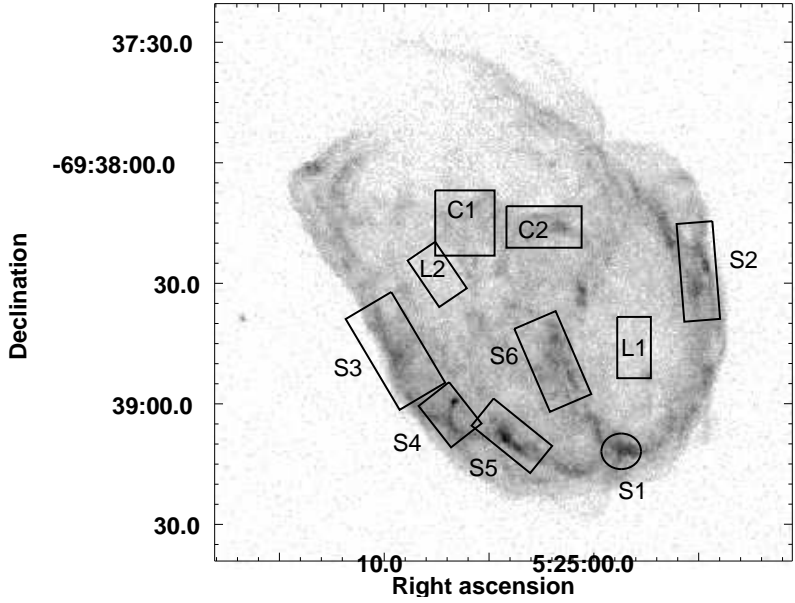}
\caption{Raw ACIS-S image of N132D in square-root brightness scales.
The labeled regions are used for spectrum extraction.}
\end{center}
\end{figure}

\begin{figure}
\begin{center}
\includegraphics[scale=0.5]{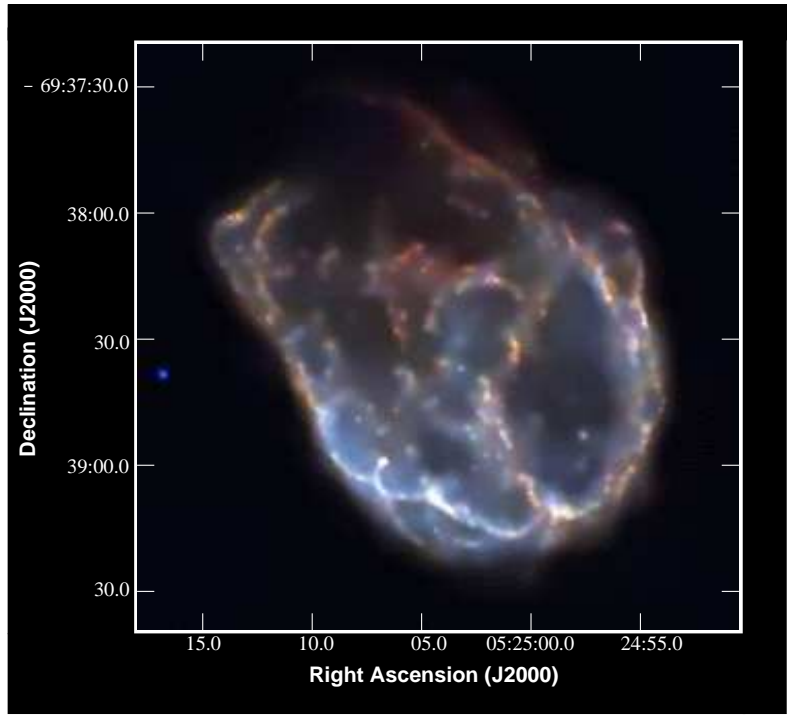}
\caption{Tricolor X-ray image of N132D: red color represents
0.3-1.0keV emission, green color represents 1.0-2.0keV emission, and
blue color represents 2.0-8.0keV emission. }
\end{center}
\end{figure}

\begin{figure}
\begin{center}
\includegraphics[scale=0.6]{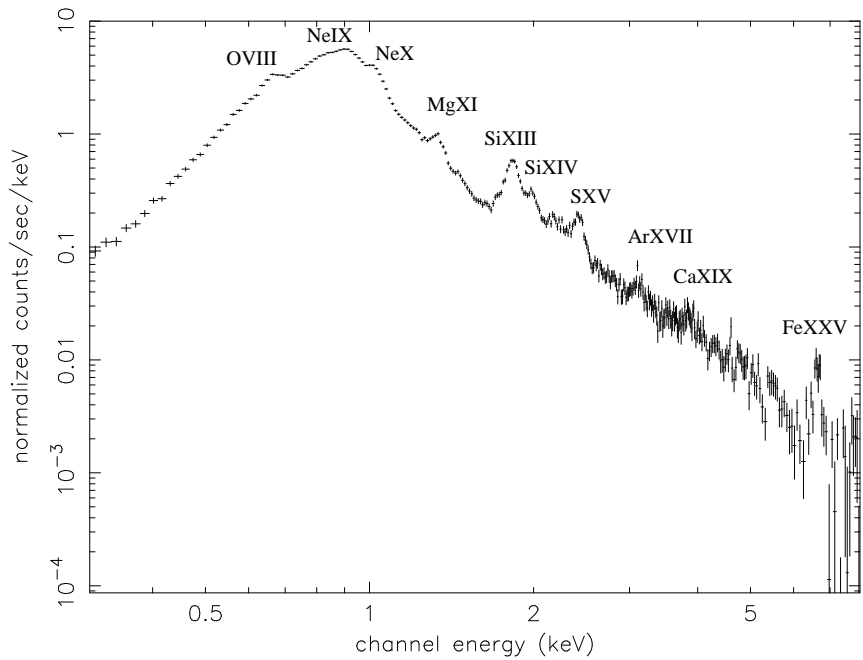}
\caption{The overall spectrum of N132D with the major emission lines
labeled}
\end{center}
\end{figure}

\begin{figure}
\centerline{
\includegraphics[scale=0.35]{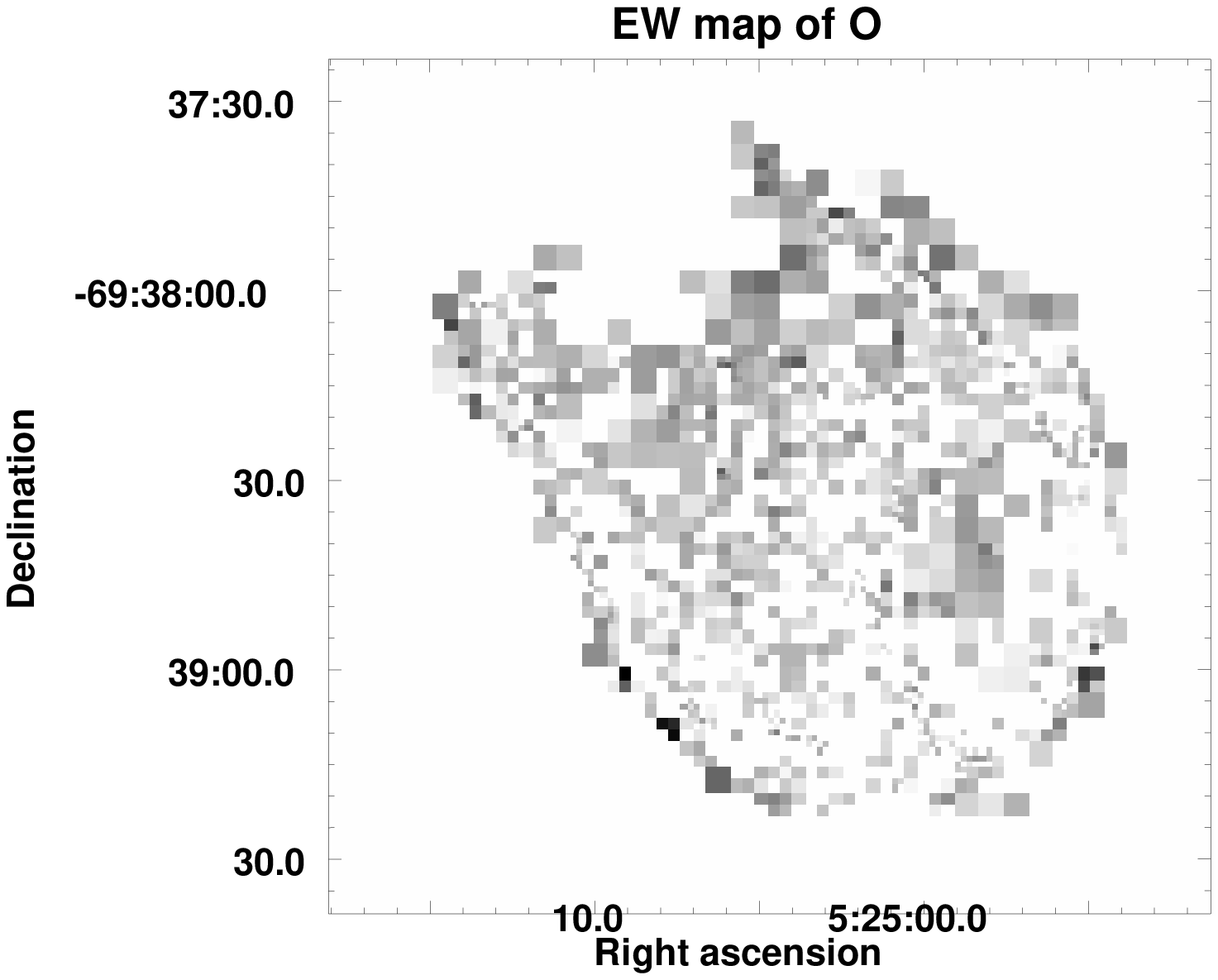}%
\hspace{0.1in}%
\includegraphics[scale=0.35]{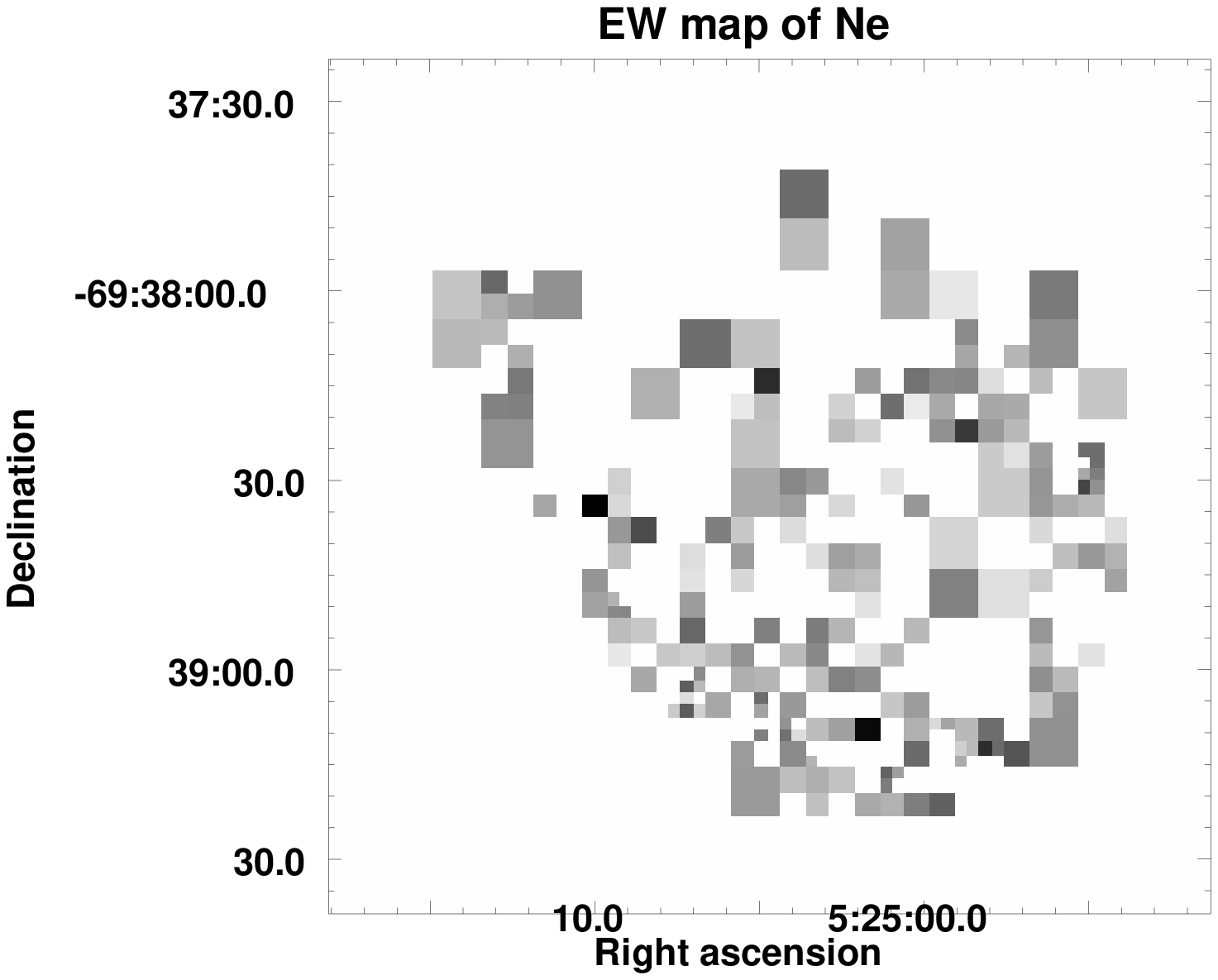}
}
\caption{Left panel: the EW map of O line at 0.65keV. Right panel:
the EW map of Ne line at 1.02keV. Both maps are in square-root
brightness scales.}
\end{figure}

\begin{figure}
\centerline{
\includegraphics[scale=0.35]{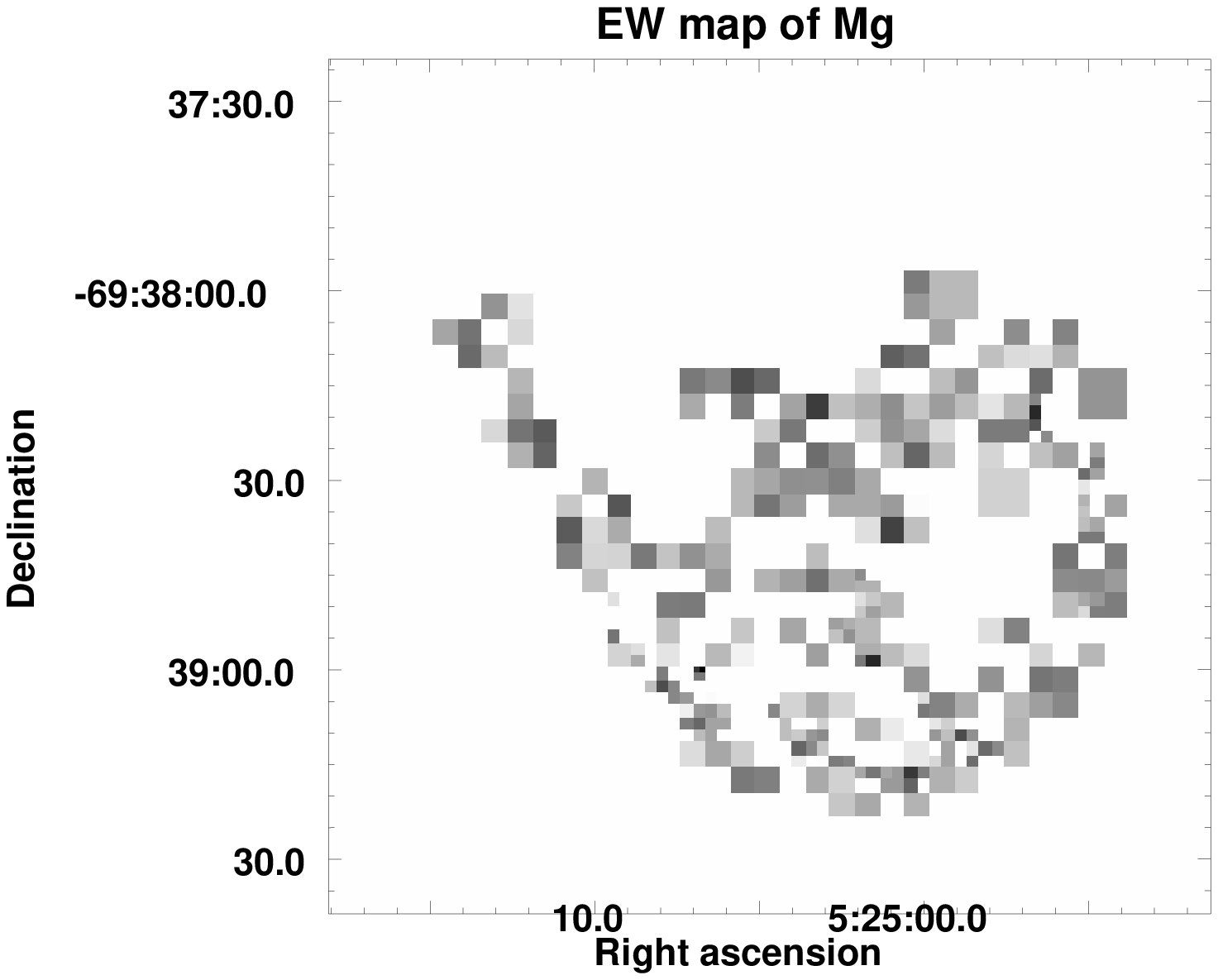}%
\hspace{0.1in}%
\includegraphics[scale=0.35]{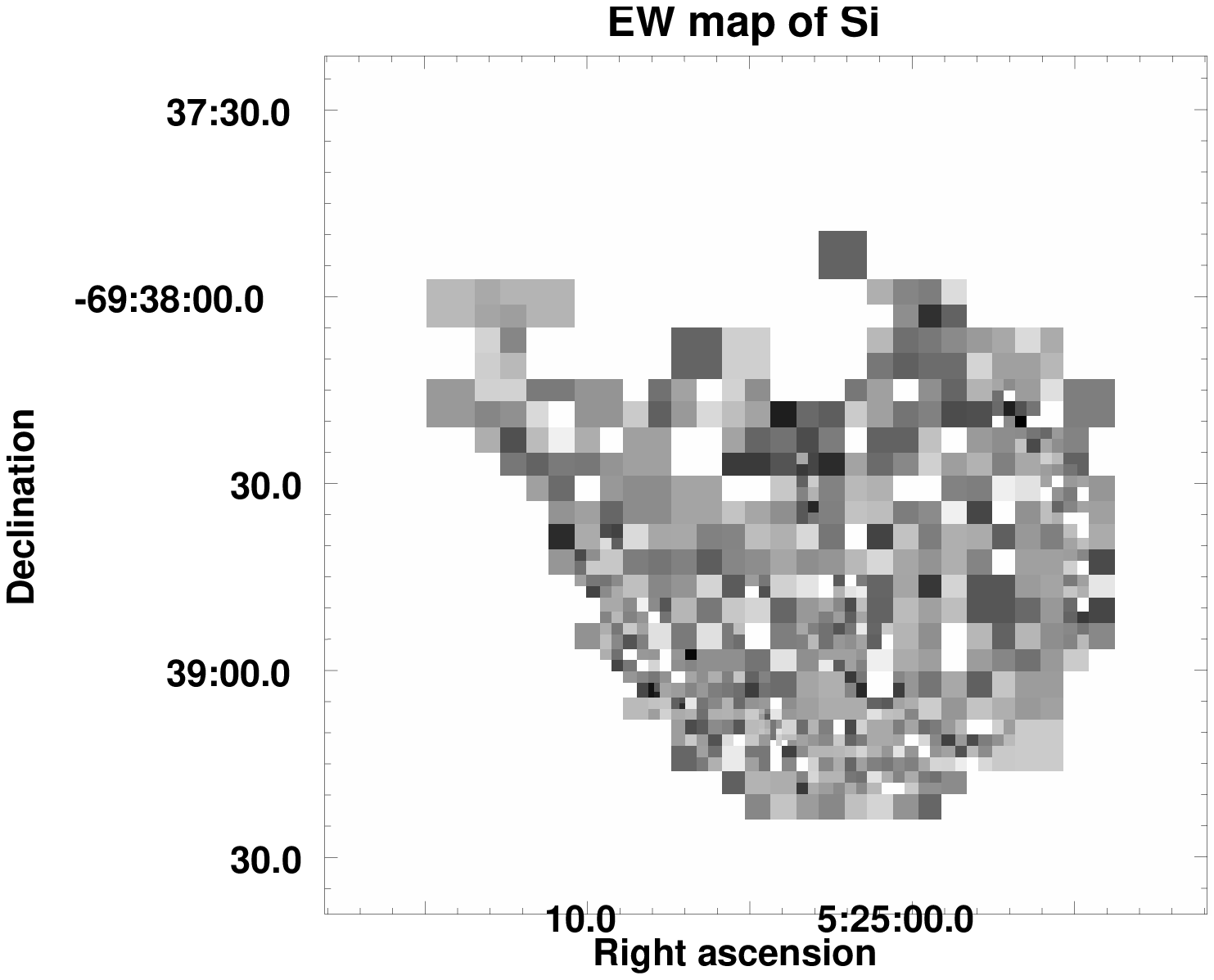}
}
\caption{Left panel: the EW map of Mg line at 1.34keV. Right panel:
the EW map of Si line at 1.87keV. Both maps are in square-root
brightness scales.}
\end{figure}

\begin{figure}
\centerline{
\includegraphics[scale=0.6]{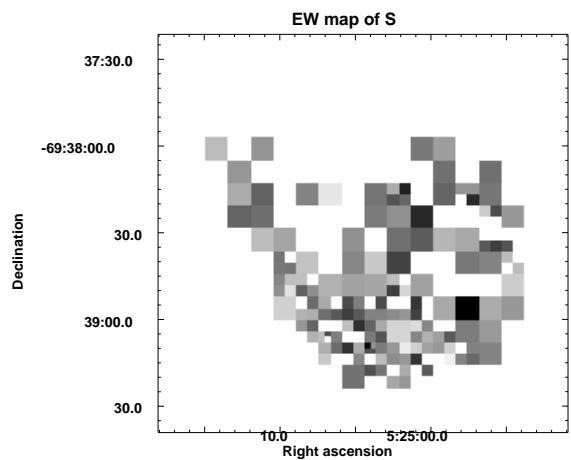}
}
\caption{ The EW map of S line at 2.45keV in square-root brightness
scales.}
\end{figure}

\subsection{Spectral Analysis}

The overall spectrum of N132D is shown in Fig.3, which shows line
features of OVIII ($\sim$0.65keV), NeIX ($\sim$0.92keV), NeX
($\sim$1.02keV), MgXI ($\sim$1.34keV), SiXIII ($\sim$1.87keV), SiXIV
($\sim$2.0keV), SXV ($\sim$2.45keV), ArXVII ($\sim$3.12keV), CaXIX
($\sim$3.86keV), and FeXXV $K_\alpha$ ($\sim$6.65keV). Due to its
complexity, at least two components are needed to fit the spectrum.
Profited from Chandra's superb angular resolution, we could perform
spatially resolved spectral analysis of N132D. We defined 10
small-scale regions (diagrammed in Fig.1) for spectral
investigation. Regions S1, S2, S3, S4, and S5 are located along the
shell, and region S6 includes a part of the bright ridge; C1 and C2
represent two bright interior regions; L1 and L2 represent two faint
interior regions. A nearby region which is not affected by photons
from N132D was selected as background. We extracted spectrum from
each region using CIAO script {\em acisspec}. All the spectra
mentioned above were regrouped to contain at least 25 net counts per
bin. The spectra were analyzed with software Xspec (ver. 11.3.1).

We fit the spectra of the small-scale regions with absorbed NEI
plane-parallel shock model {\em vpshock} and/or the power law model
{\em powerlaw}. We use the cross-sections of Morrison \& McCammon
(1983) and assume solar abundances. The spectral fit allows the
abundances of the significant line-emitting elements to vary, while
other metal abundances are set to 0.3 solar, typical for the LMC
(Russell \& Dopita, 1992). The thermal NEI model {\em vpshock}
provides the plasma temperature $kT$, the ionization timescale $n_e
t$, and the normalization parameter
\begin{math}
norm=\frac{10^{-14}}{4\pi{d}^2} \int n_en_H \mathrm{d} V
\end{math}, where $d$ is the distance to the source, $n_e$ and $n_H$
are the electron density and the hydrogen density
respectively, and $V$ is the volume of the region. Approximatively
\begin{math}
 norm=\frac{10^{-14}}{4\pi{d}^2} fn_en_H V
 \end{math},
 where $f$ is the filling factor. Volumes of the shell regions (S2-S5) are
calculated as the intersection of a sphere (the remnant) and
columns. Shell region S1 is assumed to be an oblate spheroid. The
rectangular regions C1, C2, L1, L2, and S6 are approximated as
cuboids with average line-of-sight size to be the remnant radius
$R$. Using $n_e \approx 1.2\nH$, we can estimate the density
$\nH/f^{-1/2}$ from parameter $norm$.

The spectra from interior regions C1, C2, L1, and L2 can be
described well with a single-{\em vpshock} model. Results are given
in Table 1. In the faint regions L1 and L2, the temperature is above
1keV and the abundances of heavy elements are higher than the LMC
mean while in the bright regions C1 and C2, the temperature is below
1keV and the abundances are close to the LMC mean. The
$\nH/f^{-1/2}$ values of C1 and C2 are almost twice/thrice those of
L1 and L2.

\begin{figure}
\begin{center}
\includegraphics[scale=0.5]{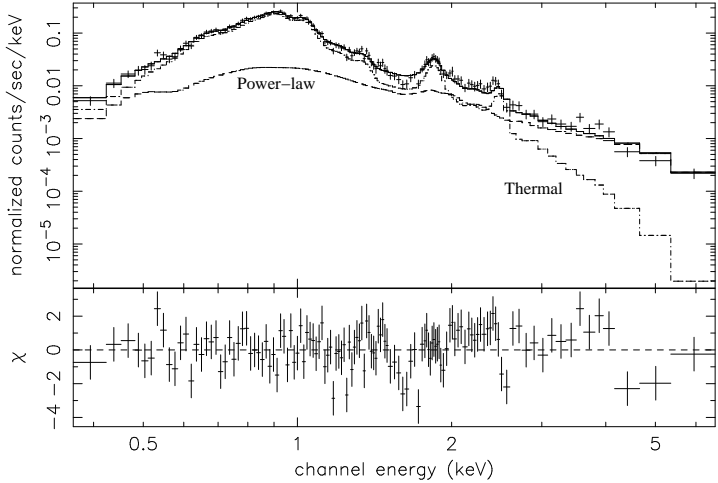}
\caption{ACIS-S spectrum of shell region S4 fitted with {\em
vpshock}+{\em powerlaw} model. The two spectral components are also
drawn separately. The lower panel shows the residuals of the data
minus the model in units of $\sigma$.}
\end{center}
\end{figure}

For the spectra of regions S1-S6, one component model is not enough
to get an acceptable fit (spectrum of S4 for example in Fig.7). We
found that there is always a hard energy tail above 2keV band which
can not be fit well with only one thermal model. Therefore we
tentatively fit these spectra with two components.

As the first possibility, the complex spectra may include two
thermal components of different temperatures. We tried to fit the
spectra with double-{\em vpshock} model. Abundances of O, Si, S, and
Fe of the hot components were thawed. The $\chi^{2}$ value is
acceptable and the results are given in Table 2. The temperatures of
the cool components are around 0.5-0.6keV while the temperatures of
the hot component are around 1.5keV. Some element abundances in S4
and S5 are slightly overabundant. The parameters derived from the
ridge-like region S6 are similar to those derived from the shell
regions except for the element abundances which are close to the
average abundance of the LMC. The ionization timescale $n_e t$ of
the cool components are high ($\geq 10^{12} cm^{-3}s$) in all
regions (S1-S6) except S4. Parameter $\nH/f^{-1/2}$ of the cool
components is almost twice/thrice those of hot components. We also
tried the two-component model with equal abundances. The
temperatures and ionization timescales are similar to those we
present here. The abundances of the "equal" model are similar to
those of the hot component in the presented model within their
errors. The problems of the two thermal components models will be
discussed in section 3.2.

Another possibility is that the spectra contain a non-thermal
component in addition to a thermal one. We fit the spectra with {\em
vpshock} + {\em powerlaw} model (spectrum of S4 for example in
Fig.7) and tabulate the results in Table 3. The photon indices are
between 2.3 and 3.4. The temperatures of the thermal components are
around 0.6keV. In the shell regions (S1-S5), some element abundances
in some regions especially S4 and S5 are slightly overabundant. The
parameters of S6 are similar to those of the shell regions again
except for the element abundances which are close to the LMC mean.

\begin{table}
\begin{center}
\caption{VPSHOCK fitting results with 90\% confidence ranges
  and estimates of the gas density for interior regions}
\newsavebox{\tablebox}
\begin{lrbox}{\tablebox}

\begin{tabular}{c|cccc}
\hline\hline Region & $L1$ & $L2$ & $C1$ & $C2$\\
\hline
$\NH$($10^{21}\cm^{-2}$)....... & $1.9^{+1.3}_{-1.0}$ & $1.3^{+1.1}_{-0.9}$ & $0.8^{+0.4}_{-0.3}$ & $0.9^{+0.3}_{-0.3}$  \\

 $kT(\keV)..............$ & $1.65^{+1.13}_{-0.44}$ & $1.20^{+0.58}_{-0.32}$ & $0.79^{+0.11}_{-0.10}$ & $0.83^{+0.15}_{-0.10}$ \\

 $norm$($10^{-4}$).......... & $2.10^{+1.20}_{-2.10}$ & $2.48^{+1.74}_{-1.17}$ & $18.55^{+6.31}_{-4.58}$ & $21.52^{+4.40}_{-5.87}$\\

$n_e t$($10^{11}\cm^{-3}\,{\rm s}$).... & $1.10^{+0.90}_{-0.43}$ & $1.23^{+1.21}_{-0.51}$ & $1.02^{+0.55}_{-0.30}$ & $1.09^{+0.62}_{-0.39}$ \\

 $O...........................$ & $1.56^{+2.33}_{-0.80}$ & $0.82^{+0.69}_{-0.43}$  & $0.37^{+0.09}_{-0.06}$ & $0.26^{+0.07}_{-0.10}$\\

$Ne.........................$  & $0.3(frozen)$ & $0.3(frozen)$ & $0.53^{+0.10}_{-0.08}$ & $0.35^{+0.11}_{-0.07}$ \\

 $Si..........................$ & $0.93^{+1.75}_{-0.39}$ & $0.42^{+0.38}_{-0.30}$ & $0.26^{+0.12}_{-0.11}$ & $0.28^{+0.13}_{-0.08}$  \\

$S...........................$  & $0.3(frozen)$ & $0.3(frozen)$ & $0.3(frozen)$ & $1.02^{+0.55}_{-0.40}$ \\

 $Fe.........................$ & $1.20^{+1.59}_{-0.50}$ & $0.72^{+0.80}_{-0.33}$ & $0.23^{+0.08}_{-0.05}$ & $0.36^{+0.11}_{-0.03}$ \\

 $net\ count\ rate$\\
 $(10^{-2}counts\s^{-1}).... $ & $1.81\pm0.04$ & $1.48\pm0.04$ & $7.00\pm0.09$ & $8.02\pm0.09$ \\

 $\chi^{2}/{\rm d.o.f.}...............$ & $1.24$ & $1.12$ & $1.10$  & $1.30$  \\

$F^{(0)}(0.3$-$8\keV)$ \\

$(10^{-11}\ergs\cm^{-2}\ps)$ & $0.168$ & $0.120$ &
$0.579$ & $0.627$ \\

$V (10^{57}\cm^{3})..........$ & $2.51 $ & $2.13 $ & $4.71$ & $3.80 $ \\

$\nH/f^{-1/2}(\cm^{-3})...$ & $1.44 $ & $1.70$ & $3.12$ & $3.74$
\\

\hline
\end{tabular}
\end{lrbox}

\scalebox{0.8}{\usebox{\tablebox}}
\end{center}
\end{table}

\begin{table}
\begin{center}
\caption{VPSHOCK+VPSHOCK fitting results with 90\% confidence ranges
  and estimates of the gas density for regions S1-S6}

\begin{lrbox}{\tablebox}

\begin{tabular}{c|cccccc}
\hline\hline Region & $S1$ & $S2$ & $S3$ & $S4$ & $S5$ & $S6$ \\
\hline $\NH$($10^{21}\cm^{-2}$)....... & $1.7^{+0.4}_{-1.0}$ &
$1.5^{+0.5}_{-0.5}$ & $1.1^{+0.2}_{-0.2}$  &
 $1.4^{+0.3}_{-0.3}$ & $1.5^{+0.3}_{-0.2}$& $1.9^{+0.7}_{-0.6}$ \\

$net\ count\ rate$\\
 $(10^{-2}counts\s^{-1}).... $ & $8.47\pm0.09$ & $14.73\pm0.13$ & $18.87\pm0.14$ & $12.64\pm0.12$ & $15.45\pm0.13$ & $17.16\pm0.14$ \\

 $\chi^{2}/{\rm d.o.f.}...............$ & $1.14$ & $1.60$ & $1.55$  & $1.38$ &
 $1.03$& $1.49$ \\

$F^{(0)}(0.3$-$8\keV)$ \\

$(10^{-11}\ergs\cm^{-2}\ps)$ & $0.729$ &
$1.25$ & $1.22$  & $0.904$ & $1.12$& $1.58$ \\

$V (10^{57}\cm^{3})...........$  & $0.19$ & $4.64 $ & $9.81 $   & $5.37 $ & $3.19 $& $ 4.94$\\

\hline

Cool Component \\

 $kT_{c}(\keV)..............$ & $0.53^{+0.05}_{-0.13}$ & $0.42^{+0.11}_{-0.03}$ & $0.63^{+0.03}_{-0.03}$ &
 $0.62^{+0.06}_{-0.05}$ & $0.59^{+0.04}_{-0.03}$ & $0.41^{+0.14}_{-0.07}$ \\

 $norm$($10^{-4}$).......... &
$40.22^{+10.88}_{-10.87} $ & $71.83^{+22.75}_{-21.34} $ &
$57.94^{+4.31}_{-10.55} $ &
 $37.87^{+6.08}_{-6.60}$ & $53.92^{+4.77}_{-5.49} $ & $86.86^{+73.74}_{-26.35}$ \\

$n_e t$($10^{11}\cm^{-3}\,{\rm s}$).... & $7.41(>5.03) $ &
$11.0(>9.32) $
& $500(>391)$  & $6.39^{+7.72}_{-2.37}$& $498(>423)$ & $83.2(>25.2)$ \\

$\nH/f^{-1/2}(\cm^{-3})....$ & $ 22.78$ & $ 6.19$ & $ 3.82$ & $
4.17$ &
$ 6.47$& $6.59$ \\

\hline

Hot Component \\

 $kT_{h}(\keV)...............$ & $1.65^{+1.19}_{-0.55}$ & $1.36^{+0.24}_{-0.15}$ & $1.59^{+0.23}_{-0.21}$  &
 $1.73^{+0.36}_{-0.24}$ & $1.53^{+0.32}_{-0.07}$& $1.35^{+0.18}_{-0.12}$ \\

 $norm$($10^{-4}$).......... & $4.24^{+8.68}_{-2.88}$ & $13.98^{+5.02}_{-4.71}$ & $16.47^{+8.06}_{-4.49}$  &
 $10.85^{+4.39}_{-3.74}$ & $16.12^{+3.78}_{-5.99}$& $23.93^{+4.98}_{-10.10}$ \\

 $n_e t$($10^{11}\cm^{-3}\,{\rm s}$).... & $0.69^{+0.33}_{-0.22}$ & $0.97^{+0.27}_{-0.18}$ & $1.68^{+0.56}_{-0.37}$  &
 $1.78^{+0.69}_{-0.45}$ & $2.05^{+0.47}_{-0.54}$& $0.873^{+0.15}_{-0.11}$ \\

$\nH/f^{-1/2}(\cm^{-3})...$ & $ 7.40$ & $2.72$ & $2.03$  & $2.23$ &
$3.53$& $3.46$ \\

 $O...........................$ & $0.21^{+0.71}_{-0.21}$ & $0.17^{+0.23}_{-0.17}$ & $0.68^{+0.24}_{-0.28}$  & $ 0.3(frozen)$
 & $0.67^{+0.28}_{-0.20}$& $0.22^{+0.12}_{-0.22}$ \\

 $Si..........................$ & $0.62^{+0.50}_{-0.37}$ & $0.56^{+0.22}_{-0.16}$ & $0.31^{+0.08}_{-0.14}$  & $0.94^{+0.45}_{-0.27}$
 & $0.45^{+0.17}_{-0.07}$& $0.35^{+0.11}_{-0.10}$ \\

 $S...........................$ & $0.86^{+0.98}_{-0.74}$ & $0.64^{+0.32}_{-0.29}$ & $0.61^{+0.23}_{-0.23}$  & $0.92^{+0.42}_{-0.33}$
 & $0.93^{+0.32}_{-0.13}$& $0.28^{+0.20}_{-0.10}$ \\

 $Fe.........................$ & $1.00^{+1.00}_{-0.41}$ & $0.56^{+0.18}_{-0.12}$ & $0.67^{+0.17}_{-0.14}$  &
 $0.85^{+0.34}_{-0.21}$ & $0.73^{+0.37}_{-0.13}$& $0.57^{+0.13}_{-0.10}$ \\

\hline
\end{tabular}
\end{lrbox}

\scalebox{0.73}{\usebox{\tablebox}}
\end{center}
\end{table}

\begin{table}
\begin{center}
\caption{POWERLAW+VPSHOCK fitting results with 90\% confidence
ranges and estimates of the gas density for regions S1-S6}

\begin{lrbox}{\tablebox}

\begin{tabular}{c|cccccc}
\hline\hline Region & $S1$ & $S2$ & $S3$ & $S4$ & $S5$ & $S6$ \\
\hline $\NH$($10^{21}\cm^{-2}$)....... & $1.8^{+0.3}_{-0.5}$ &
$1.5^{+0.4}_{-0.4}$ & $1.7^{+0.2}_{-0.2}$  &
 $2.8^{+0.4}_{-0.2}$ & $2.0^{+0.5}_{-0.4}$& $1.5^{+0.3}_{-0.3}$ \\

$net\ count\ rate$\\
 $(10^{-2}counts\s^{-1}).... $ & $8.47\pm0.09$ & $14.73\pm0.13$ & $18.87\pm0.14$ & $12.64\pm0.12$ & $15.45\pm0.13$ & $17.16\pm0.14$ \\

 $\chi^{2}/{\rm d.o.f.}................$ & $1.15$ & $1.75$ & $1.53$  & $1.41$ &
 $1.17$& $1.65$ \\

$F^{(0)}(0.3$-$8\keV)$\\

$(10^{-11}\ergs\cm^{-2}\ps)$ & $0.757 $ & $1.27$ & $1.66
$  & $1.84$ & $1.40$& $1.34$ \\

$V (10^{57}\cm^{3})..........$ & $0.19$ & $4.64 $ & $9.81 $   & $5.37 $ & $3.19 $& $ 4.94$  \\

\hline Powerlaw component\\

 $PhoIndex..............$ & $2.57^{+0.29}_{-1.24}$ & $3.09^{+0.35}_{-0.39}$ & $3.05^{+0.31}_{-0.41}$  &
 $3.40^{+0.22}_{-0.26}$ & $2.83^{+0.45}_{-0.22}$& $2.37^{+0.45}_{-0.24}$ \\

 $norm$($10^{-4}$).......... & $0.67^{+1.22}_{-0.56}$ & $2.94^{+2.36}_{-1.78}$ & $5.53^{+3.34}_{-2.48}$  &
 $8.23^{+3.06}_{-1.28}$ & $3.51^{+3.89}_{-1.78}$& $1.52^{+1.57}_{-0.76}$ \\

\hline Thermal component\\
 $kT_{x}(\keV)...............$ & $0.60^{+0.04}_{-0.04}$ & $0.54^{+0.04}_{-0.07}$ & $0.65^{+0.02}_{-0.02}$  &
 $0.54^{+0.07}_{-0.05}$ & $0.61^{+0.02}_{-0.04}$& $0.60^{+0.02}_{-0.01}$ \\

 $norm$($10^{-4}$).......... & $43.21^{+10.51}_{-8.94}$ & $68.26^{+19.90}_{-14.89}$ & $71.15^{+13.15}_{-13.03}$  &
 $62.11^{+17.41}_{-15.47}$ & $72.94^{+12.59}_{-15.60}$& $89.93^{+6.43}_{-11.92}$ \\

 $n_e t$($10^{11}\cm^{-3}\,{\rm s}$).... & $5.91^{+10.74}_{-3.00}$ & $7.93^{+5.34}_{-2.94}$ & $9.95^{+8.58}_{-4.20}$  &
 $4.93^{+2.43}_{-1.46}$ & $20.53^{+12.94}_{-11.84}$& $15.5^{+13.15}_{-7.65}$ \\

$\nH/f^{-1/2}(\cm^{-3})...$ & $ 23.36$ & $6.03$ & $4.24$  & $5.35$ &
$7.52$& $6.71$ \\

 $O...........................$ & $0.36^{+0.24}_{-0.14}$ & $0.44^{+0.29}_{-0.13}$ & $0.73^{+0.36}_{-0.11}$  & $0.42^{+0.20}_{-0.12}$
 & $0.81^{+0.39}_{-0.26}$& $0.53^{+0.12}_{-0.16}$ \\

 $Ne.........................$ & $0.49^{+0.21}_{-0.12}$ & $0.65^{+0.19}_{-0.13}$ & $1.06^{+0.23}_{-0.18}$  & $1.18^{+0.45}_{-0.30}$ & $1.25^{+0.41}_{-0.28}$&
 $0.61^{+0.17}_{-0.14}$ \\

 $Si..........................$ & $0.35^{+0.13}_{-0.10}$ & $0.46^{+0.14}_{-0.11}$ & $0.32^{+0.11}_{-0.09}$  & $0.79^{+0.30}_{-0.20}$& $0.38^{+0.08}_{-0.05}$
 & $0.25^{+0.07}_{-0.06}$ \\

 $S...........................$ & $0.63^{+0.41}_{-0.38}$ & $0.92^{+0.41}_{-0.40}$ & $0.88^{+0.31}_{-0.29}$  & $2.17^{+0.59}_{-0.67}$& $1.31^{+0.40}_{-0.34}$
 & $0.34^{+0.21}_{-0.19}$ \\

 $Fe.........................$ & $0.37^{+0.10}_{-0.06}$ & $0.33^{+0.08}_{-0.05}$ & $0.36^{+0.11}_{-0.07}$  &
 $0.31^{+0.11}_{-0.07}$& $0.32^{+0.11}_{-0.05}$ & $0.31^{+0.05}_{-0.03}$ \\

\hline
\end{tabular}
\end{lrbox}

\scalebox{0.75}{\usebox{\tablebox}}
\end{center}
\end{table}

\section{Discussion}

\subsection{The interior regions}

For the interior regions that have been spectrally analyzed, we find
that the faint regions L1 and L2 have higher temperature (above
1keV) and overabundant heavy elements while interior bright regions
C1 and C2 have lower temperature (below 1keV) and lower abundances
close to the LMC mean. We note that parameter $\nH/f^{-1/2}$ derived
from the volume emission measure of the gas in the bright regions is
about twice/thrice those in the faint regions. The products of $\nH$
and $kT$ for these regions are around $2.4f^{-1/2}$ $cm^{-3}keV$,
suggesting that the cool dense regions may be in pressure balance
with the hot tenuous regions (given that the filling factors are
similar).

\subsection{The shell regions}
The X-ray spectra of the shell regions S1-S5 have at least two
components. Both the double-{\em vpshock} model and {\em vpshock} +
{\em powerlaw} model fit the X-ray spectra of these regions well.

In the double-{\em vpshock} case, the high ionization timescale $n_e
t$ ($\geq 10^{12} cm^{-3}s$) of the cool components indicates that
the cool gas has approached or been in the ionization equilibrium.
However, the ionization age derived for the shocked gas seems too
large. Taking S3 for example, we estimate the ionization age
$t\approx 3\times 10^{5} f^{1/2}$ $yr$. The kinematic age of N132D
is 3150 $\pm$ 200 years (Morse et al., 1995). Unless the filling
factor is as small as $10^{-4}$, the ionization age will be too old
to understand. On the other hand, the outer shell is commonly
considered to be swept-up ISM. Therefore the co-existence of two
thermal components along the shell is difficult to explain. One
possibility is that the regions we selected to obtain sufficient
counts statistics are quite large in physical extent and contain
emission from gas at varying thermodynamic states.

In the {\em vpshock}+{\em powerlaw} case, the spectra provide a
possibility for the non-thermal component to be present along the
shell. Although the theories of X-ray synchrotron emission from SNR
shell still have open questions, more and more shell-like SNRs have
been confirmed to exhibit X-ray synchrotron emission. Two shell-like
SNRs, SN 1006, and G347.3-0.5, are found to be dominated by
synchrotron X-rays with no emission line (Koyama et al., 1995;
Koyama et al., 1997; Slane et al., 1999) while some shell-like SNRs
such as Cas~A, Kepler, Tycho, and RCW86 are predominantly thermal but
contain a hard energy tail due to electron synchrotron. Like these
SNRs, N132D is young and has a shell-like morphology. It is thus
interesting to find a hard energy tail in the shell of N132D. The
presence of X-ray synchrotron may be helpful in understanding the
positional coincidence of the X-ray shell with the radio one (Dickel
\& Milne, 1995). The best-fit photon indices we derived are between 2.3 and
3.4 which are consistent with the typical value (2.5-3) for diffuse
shock acceleration in those shell-like SNRs (Ballet, 2006). If the
non-thermal component could be further confirmed, N132D would be
listed as another X-ray synchrotron SNR. The slightly enhanced
element abundances in some portions of the shell might be ascribed
to the dust destruction by sputtering.

\subsection{The ridge}

The ridge region S6 shows some consistency with the shell regions
S1-S5. In morphology, the ridge connects the shell at the south. In
spectra, the parameters of S6 derived from the two models are
similar to those of S1-S5 except for some lower element abundances.
The spitzer images show that both the ridge and the shell have
strong $24\mu m$ emission which is from swept-up, shock-heated dust
grains (Tappe et al., 2006). So we suggest that the ridge region S6
may be the projection of a part of shell in which the ISM is the
primary ingredient.

\section{Conclusion}
 Using the zeroth-order data of Chandra HETG observation, we tentatively
 performed a spatially-resolved X-ray spectroscopy study of SNR N132D. The EW
maps of O, Ne, Mg, Si, and S are generated. The faint interior
regions have lower density, higher temperature above 1keV than that
of bright interior regions. The X-ray spectra along the shell can be
phenomenally described with either a double-{\em vpshock} model or a
{\em vpshock} + {\em powerlaw} model. For the double-{\em vpshock}
case, the low temperature component has a unreasonable high
ionization timescale, and two thermal components along the shell as
swept-up ISM are difficult to understand. For the {\em vpshock} +
{\em powerlaw} model, Ne and S are slightly overabundant in some
regions of the shell, and the best-fit photon indices of the non-thermal
component are between 2.3 and 3.4, consistent with diffuse shock
acceleration. If the non-thermal component is true, N132D would be
listed as another TeV particle acceleration laboratory.

We thank the two anonymous referees for a thorough review of this
work and for useful comments. We also thank Yang Su and Jiang-Tao Li
for technical assistance. Y.C.\ acknowledges support from NSFC
grants 10221001 and 10673003.


\begin{thebibliography}{}



\small


\harvarditem{Ballet}{2006}{Ballet06} Ballet, J. X-ray synchrotron
emission from supernova remnants. AdSpR. 37, 1902, 2006.
%


\harvarditem{Behar et al.}{2001}{Behar01} Behar, E., Rasmussen, A.
P., Griffiths, R. G., Dennerl, K., Audard, M., Aschenbach, B.,
Brinkman, A. C. High-resolution X-ray spectroscopy and imaging of
supernova remnant N132D. Astron. Astrophys. 365, L242--L247, 2001.
%

\harvarditem{Blair et al.}{1994}{Blair94} Blair, W. P., Raymond, J.
C., Long, K. S. IUE spectra and optical imaging of the oxygen-rich
supernova remnant N132D. Astrophys. J. 423, 334--343, 1994.
%

\harvarditem{Blair et al.}{2000}{Blair00} Blair, W. P., Morse, J.
A., Raymond, J. C., Kirshner, R. P., Hughes, J. P., Dopita, M. A.,
Sutherland, R. S., Long, K. S., Winkler, P. F. Hubble Space
Telescope Observations of Oxygen-rich Supernova Remnants in the
Magellanic Clouds. II. Elemental Abundances in N132D and 1E
0102.2-7219. Astrophys. J. 537, 667--689, 2000.
%

\harvarditem{Canizares et al.}{2001}{Canizares01}
 Canizares, C. R., Flanagan, K. A., Davis, D. S., Dewey, D., Houck, J.
 C. High Resolution Spectroscopy of Two Oxygen-Rich SNRs with the Chandra
 HETG. ASPC. 234, 173--180, 2001.

%


\harvarditem{Chen et al.}{2003}{Chen03} Chen, Y., Zhang, F.,
Williams, R. M., Wang, Q. D. Supernova Remnant Crossing a Density
Jump: A Thin-Shell Model. Astrophys. J. 595, 227, 2003.

%



\harvarditem{Danziger \& Dennefeld}{1976}{Danziger76}
 Danziger, I. J. \& Dennefeld, M. Supernova ejecta in the Large Magellanic Cloud. Astrophys. J. 207, 394--407, 1976.
%

\harvarditem{Dickel \& Milne}{1995}{Dickel95} Dickel, J. R. \&
Milne, D. K. Radio properties of three young supernova remnants in
the Large Magellanic
 Cloud. Astron. J. 109, 200--208, 1995.

%




\harvarditem{Favata et al.}{1997}{Favata97} Favata, F., Vink, J.,
Parmar, A. N., Kaastra, J. S. Mineo, T. BeppoSAX LECS/MECS X-ray
spectroscopy of the young supernova remnant N132D. Astron.
Astrophys. 324, L45--L48, 1997.
%


\harvarditem{Favata et al.}{1998}{Favata98} Favata, F., Vink, J.,
Parmar, A. N., Kaastra, J., Mineo, T. Erratum: BeppoSAX LECS/MECS
X-ray spectroscopy of the young supernova remnant N132D. Astron.
Astrophys. 340, 626, 1998.
%

\harvarditem{Hughes}{1987}{Hughes87} Hughes, J. P. X-ray studies of
the supernova remnant N132D. I - Morphology. Astrophys. J. 314,
103--110, 1987.

%
\harvarditem{Hughes et al.}{1998}{Hughes98} Hughes, J. P., Hayashi,
I., Koyama, K. ASCA X-Ray Spectroscopy of Large Magellanic Cloud
Supernova Remnants and the Metal Abundances of the Large Magellanic
Cloud. Astrophys. J. 505, 732--748, 1998.


 \harvarditem{Hwang et al.}{1993}{Hwang93} Hwang, U.,
Hughes, J. P., Canizares, C. R., Markert, T. H. High-resolution
X-ray spectroscopy of the supernova remnant N132D. Astrophys. J.
414, 219--229, 1993.
%

\harvarditem{Hwang et al.}{2000}{Hwang00} Hwang, U., Holt, S. S.,
Petre, R. Mapping the X-Ray-emitting Ejecta in Cassiopeia A with
Chandra. Astrophys. J. 537, L119--L122, 2000.


\harvarditem{Koyama et al.}{1995}{Koyama95} Koyama, K., Petre, R.,
Gotthelf, E. V., Hwang, U., Matsuura, M., Ozaki, M., Holt, S. S.
Evidence for Shock Acceleration of High-Energy Electrons in the
Supernova Remnant SN:1006. Nature. 378, 255--258, 1995.
%


\harvarditem{Koyama et al.}{1997}{Koyama97} Koyama, K., Kinugasa,
K., Matsuzaki, K., Nishiuchi, M., Sugizaki, M., Torii, K., Yamauchi,
S., Aschenbach, B. Discovery of Non-Thermal X-Rays from the
Northwest Shell of the New SNR RX J1713.7-3946: The Second SN 1006?.
PASJ. 49, L7--L11, 1997.
%


\harvarditem{Lasker}{1978}{Lasker78} Lasker, B. M. Studies of N132D,
a supernova remnant similar to Cassiopeia A in the large Magellanic
cloud. Astrophys. J. 223, 109--121, 1978.

%


\harvarditem{Mathewson et al.}{1983}{Mathewson83} Mathewson, D. S.,
Ford, V. L., Dopita, M. A., Tuohy, I. R., Long, K. S., Helfand, D.
J. Supernova remnants in the Magellanic Clouds. Astrophys. J.S. 51,
345--355, 1983.
%



\harvarditem{Morrison \& McCammon}{1983}{Morrison83} Morrison, R. \&
McCammon, D. Interstellar photoelectric absorption cross sections,
0.03-10 keV. Astrophys. J. 270, 119--122, 1983.

%


\harvarditem{Morse et al.}{1995}{Morse95} Morse, J. A., Winkler, P.
F., Kirshner, R. P. Spatially Resolved Kinematics and Longslit
Spectroscopy of the Young, Oxygen-Rich Supernova Remnant N132D in
the Large Magellanic Cloud. Astrono. J. 109, 2104--2120, 1995.

%
\harvarditem{Morse et al.}{1996}{Morse96} Morse, J. A., Blair, W.
P., Dopita, M. A., Hughes, J. P., Kirshner, R. P., Long, K. S.,
Raymond, J. C., Sutherland, R. S., Winkler, P. F. Hubble Space
Telescope Observations of Oxygen-Rich Supernova Remnants in the
Magellanic Cloud. I. Narrow-Band Imaging of N132D in the LMC.
Astrono. J. 112, 509--533, 1996.
%








\harvarditem{Russell \& Dopita}{1992}{Russell92} Russell, S. C.,
Dopita, M. A. Abundances of the heavy elements in the Magellanic
Clouds. III - Interpretation of results. Astrophys. J. 384,
508--522, 1992.
%




\harvarditem{Slane et al.}{1999}{Slane99} Slane, P., Gaensler, B.,
Dame, T. M., Hughes, J. P., Plucinsky, P. P., Green, A. Nonthermal
X-Ray Emission from the Shell-Type Supernova Remnant G347.3-0.5.
Astrophys. J. 525, 357--367, 1999.
%






\harvarditem{Tappe et al.}{2006}{Tappe06} Tappe, A., Rho, J., Reach,
W. T. Shock Processing Of Interstellar Dust And Polycyclic Aromatic
Hydrocarbons In The Supernova Remnant N132D. Astrophys. J. 653, 267,
2006.

%





\end{thebibliography}
\end{document}